\documentclass{article}

\usepackage{amsmath}
\usepackage{amssymb}
\usepackage{url}
\usepackage{threeparttable}
\usepackage[utf8]{inputenc}
\usepackage{authblk}
\usepackage{hyperref}
\usepackage{graphicx}
\usepackage[authoryear]{natbib}
\usepackage[a4paper, left=3cm, right=3cm, top=3cm, bottom=3cm]{geometry}

\title{Integrating simultaneous Interfacial Shear Rheology with Neutron Reflectometry for Structural and Dynamic Analysis of fluid Interfacial Systems}

\author[1]{Pablo S\'anchez-Puga\thanks{Corresponding author. Present address: Department of Physics and Astronomy, Uppsala University, Box 516, Uppsala S-751 20, Sweden. Email: \href{mailto:sanchez-puga@ill.fr}{sanchez-puga@ill.fr}; \href{mailto:pablo.sanchez-puga@physics.uu.se}{pablo.sanchez-puga@physics.uu.se}. ORCID: \href{https://orcid.org/0000-0002-1000-7527}{0000-0002-1000-7527}}}
\author[2]{Javier Tajuelo\thanks{Email: \href{mailto:jtajuelo@ccia.uned.es}{jtajuelo@ccia.uned.es}. ORCID: \href{https://orcid.org/0000-0002-4085-7610}{0000-0002-4085-7610}}}
\author[1]{Javier Carrascosa-Tejedor\thanks{Email: \href{mailto:carrascosa-tejedor@ill.fr}{carrascosa-tejedor@ill.fr}. ORCID: \href{https://orcid.org/0000-0003-0385-2383}{0000-0003-0385-2383}}}
\author[3]{Miguel \'Angel Rubio\thanks{Email: \href{mailto:mar@fisfun.uned.es}{mar@fisfun.uned.es}. ORCID: \href{https://orcid.org/0000-0002-4210-0443}{0000-0002-4210-0443}}}
\author[1]{Philipp Gutfreund\thanks{Email: \href{mailto:gutfreund@ill.fr}{gutfreund@ill.fr}. ORCID: \href{https://orcid.org/0000-0002-7412-8571}{0000-0002-7412-8571}}}
\author[4,5]{Armando Maestro\thanks{Email: \href{mailto:armando.maestro@ehu.eus}{armando.maestro@ehu.eus}. ORCID: \href{https://orcid.org/0000-0002-7791-8130}{0000-0002-7791-8130}}}

\affil[1]{Institut Laue-Langevin, 38042, Grenoble, France}
\affil[2]{Departamento de F\'isica Interdisciplinar, Facultad de Ciencias, Universidad Nacional de Educaci\'on a Distancia (UNED), 28232, Las Rozas, Spain}
\affil[3]{Departamento de F\'isica Fundamental, Facultad de Ciencias, Universidad Nacional de Educaci\'on a Distancia (UNED), 28232, Las Rozas, Spain}
\affil[4]{Centro de F\'isica de Materiales (CFM-MPC), CSIC-EHU, Paseo Manuel de Lardizabal 5, 20018, Donostia-San Sebasti\'an, Spain}
\affil[5]{IKERBASQUE-Basque Foundation for Science, Bilbao, Spain}

\date{}
\begin{document} 

% This is the Author Accepted Manuscript of an article accepted for 
% publication in Journal of Applied Crystallography, before journal formatting.
% See the final version at https://journals.iucr.org/j/

\maketitle 

\noindent\textbf{Synopsis:} We present a new setup at the FIGARO horizontal neutron reflectometer enabling simultaneous measurement of the structural and mechanical properties of liquid interfaces, combining neutron reflectometry with interfacial shear rheology.

\begin{abstract}
The study of the structure and mechanical properties of complex fluid interfaces has attracted growing interest in recent decades due to its fundamental relevance to biological systems, drug development, and industrial applications. A central challenge in this field is establishing a direct link between the macroscopic mechanical response of an interface and its underlying molecular-scale structural evolution. To address this, we present an integrated experimental approach that combines interfacial rheology with neutron reflectometry, enabling simultaneous measurement of dynamical and structural properties on the same sample. This strategy eliminates the uncertainties inherent in comparing separately prepared samples and, more importantly, provides direct mechanistic insight into phenomena such as film formation, phase transitions, and kinetic processes. We validated the methodology using Langmuir monolayers of the saturated phospholipid 1,2-dipalmitoyl-sn-glycero-3-phosphocholine (DPPC) at the air/water interface. The measurements were performed using a newly developed sample environment for the FIGARO horizontal neutron reflectometer at the Institut Laue-Langevin, which integrates a double wall-ring (DWR) interfacial shear rheometer compatible with commercial rotational rheometers. This innovative setup paves the way for broad application to complex interfacial systems—including polymers, biomembranes, and multi-layer films—where coupled structural–mechanical insight is critical.
\end{abstract}


\section{Introduction}
\label{sec:intro}

Fluid interfaces are found in living systems and various technological processes. Currently, significant scientific effort is being devoted to exploring the potential of supramolecular assemblies composed of large, multifunctional colloidal nano-objects, which encompass amphiphilic molecules, macromolecules, polymers, and organic and metallic nanoparticles \cite{Guzman2016,Maestro2019, Guzman2022}. These interfaces often exhibit complex structural organisation, displaying a non-linear response to mechanical deformations \cite{Fuller2012}. The interrelationship between the structural and dynamical characteristics of these complex fluid interfaces is crucial in numerous natural and technological processes \cite{Lopez2020}. Increasing our knowledge of these phenomena is essential for understanding the fundamentals of various biological processes, the development of new drugs, consumer products, or other industrial applications \cite{Maestro2019}. Examples of such complex interfaces are the phospholipd bilayer that composes cell membranes together with the inclusion of other chemical compounds (cholesterol, proteins, fatty acids, etc.) \cite{Waldie2020}, the pulmonary surfactant \cite{Collada2026}, or the tear film. In addition, interfaces are inherent to many products in the food, personal care, and biotherapeutic sectors, where emulsions and foams are ubiquitous \cite{Maestro2014, Maestro2018}.

Several scattering techniques have been developed so far to address interfacial molecular structures \cite{Kaganer1999}. In particular, neutron reflectometry (NR) and X-ray reflectometry (XRR) \cite{Braun2017, Lu2000, Maestro2021}, together with grazing incidence X-ray diffraction (GIXD) \cite{Gerber2006, Daillant2008, Krafft2001}, have been successfully used to reveal both the in-plane and out-of-plane molecular structure of surface films. XRR offers the advantage of covering a broad range of length scales, enabling high-resolution measurements, while GIXD allows the extraction of fine details of the crystallographic morphology at the molecular scale. However, their high-flux density and electronic interaction can be detrimental to various soft matter systems. In contrast, NR, while not covering the same momentum transfer range as XRR, provides the unique advantage of allowing scattering contrast variation, achieved through sample deuteration or the use of mixtures of light and heavy water, in the case of aqueous solutions. Furthermore, NR is considerably less invasive compared to XRR, making it more suitable for the study of delicate soft matter samples.

In addition to large-scale facility based techniques, other approaches have been used to characterise the microstructure in the plane of fluid interfaces such as atomic force microscopy (AFM) \cite{Gonzalez-Martinez2019a, Gonzalez-Martinez2019b}, ellipsometry \cite{Nestler2017,Maestro2015,Ducharme2001, Maestro2021}, fluorescence microscopy \cite{Vutukuri2020, Beltramo2016} or Brewster angle microscopy (BAM) \cite{Riviere1994, Carrascosa-Tejedor2022, Carrascosa-Tejedor2023}, which are of great interest as complementary techniques to observe the formation of domains above the micrometre scale. 

With regard to interfacial rheology, several instruments have been designed for the study of the dynamical behaviour of interfaces. For shear deformation, examples include several Interfacial Shear Rheometers (ISRs) such as the magnetic needle ISR \cite{Brooks1999, Tajuelo2015, Tajuelo2016}, the microbutton \cite{Zell2016}, and specially designed fixtures for commercial rotational rheometers such as the conical bob \cite{Erni2003, Tajuelo2018, Sanchez-Puga2018} and the Double Wall-Ring (DWR) \cite{Vandebril2010}. For dilatational measurements, achieving pure deformations in experiments is rather difficult. Indeed, barrier compression techniques in rectangular Pockels-Langmuir troughs impose mixed deformations because they induce changes in both the form and the area of the interface. Nevertheless, procedures have been devised to obtain rheological information from such measurements \cite{Petkov2000}. Only recently have experimental procedures to induce pure dilatational deformations been implemented in the so-called radial trough, in its original form \cite{Pepicelli2017} or later versions \cite{Kale2021,Huang2025}, and Quadrotrough configurations \cite{Tein2022,Ashkenazi2024}.

Although some of the studies mentioned above on interfacial rheology include observations of the interface using microscopy, several other works have specifically focused on combining optical and mechanical techniques to directly link mechanical and structural properties at microscale in the context of particle-stabilized systems \cite{Keim2013, Barman2016, Alicke2023}. These approaches provide valuable insight at the micrometre scale, while simultaneous access to molecular-level structural information can be achieved using scattering-based techniques.

We remark that, unless the interfacial flow is largely dominated by the interfacial drag on the probe, a circumstance that cannot be known \textit{a priori}, adequately accounting for inertia effects and drag of the bulk phases, and properly separating elastic and viscous contributions of the interface response is mandatory. Such tasks can be conveniently carried out in the case of shear rheometry measurements using flow field-based data analysis schemes (FFBDA) \cite{Reynaert2008,Vandebril2010,Verwijlen2011,Tajuelo2015,Tajuelo2016,Tajuelo2018,Sanchez2021}.

Integration of molecular-level structural data with dynamical measurements from interfacial rheometry offers a comprehensive picture of interfacial systems and facilitates a more reliable interpretation of their mechanical response. This is the reason for the growing interest in the combination of NR and Interfacial Rheology (IR) experimental data \cite{Tein2022,Thompson2025}. From a practical point of view, the high sensitivity of interfacial systems to temperature, evaporation, and other experimental conditions, especially if working near phase transitions or meta-stable states, makes it challenging to rigorously compare structural and rheological measurements performed on separate samples. Hence, the availability of experimental facilities that offer the possibility of performing simultaneous NR and IR measurements \cite{Novaes2025} is of primary importance for the study of interfacial systems.

In this work, we present an experimental setup that allows the integration of an interfacial shear rheometry system for simultaneous measurements on the FIGARO \cite{Campbell2011} horizontal reflectometer at the Institut Laue-Langevin (ILL), which is now available as an integrated sample environment for the interfacial science community. To our knowledge, there is no neutron or synchrotron facility worldwide that offers as a standard feature the possibility of making simultaneous structural and high-sensitivity dynamical shear measurements on the same fluid-fluid interface sample. 

This paper is organised in two main sections. Section I provides a description of the DWR design and data acquisition methodology, while Section II focuses on the device's performance, including experimental validation using a 1,2-dipalmitoyl-sn-glycero-3-phosphocholine (DPPC) monolayer at the air/water interface, whose structure and dynamical behaviour have been previously reported in separate experiments \cite{Hermans2014, Campbell2018a, Carrascosa2020}.

\section{Design, operation, and data analysis}
\subsection{Experimental setup}

\begin{figure}
    \centering
    \includegraphics[width=.7\textwidth]{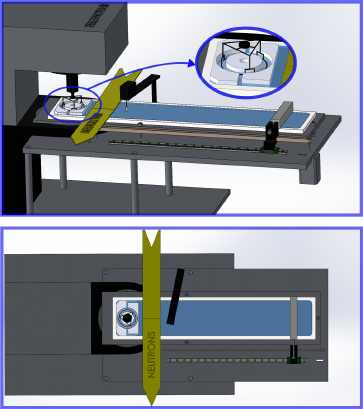}
    \caption{Sketch of the DWR on FIGARO setup. (a) Perspective of the whole ensemble attached to the rheometer frame including the support table. (Inset: Detail of the DWR ensemble). (b) Top view with a half-height cutting plane showing the disposition of the interfacial rheology measurement system, incident neutron beam and footprint, interfacial pressure balance position, and barrier travel range.}
    \label{fig:setup_panel}
\end{figure}

{\bf Langmuir trough and support system.} The mechanical setup designed for the FIGARO beamline at the ILL, as shown in Figure \ref{fig:setup_panel}, includes a custom-designed Langmuir trough featuring a single moving barrier, which has been integrated with a commercial Anton Paar MCR702e Space rheometer through an auxiliary support table system. The trough, machined on a single poly-tetrafluoroethylene plate and mounted on an aluminium plate, incorporates a copper tube circuit at its base that can be connected to a thermostatic bath, ensuring accurate temperature regulation of the sample. 

The mobile barrier, made in POM, traverses the Langmuir trough's top via a carriage mechanism on a rail, driven by a stepper motor attached to a toothed belt. The trough itself measures $101$ mm in width, and the barrier's $450$ mm travel range allows for compression ratios slightly above 5. An interfacial pressure sensor/microbalance (Kibron\textsuperscript{TM}) equipped with a Wilhelmy plate $4$ mm wide is used for the measurement and control of the interfacial pressure.

\begin{figure}
    \centering
    \includegraphics[width = .7\textwidth]{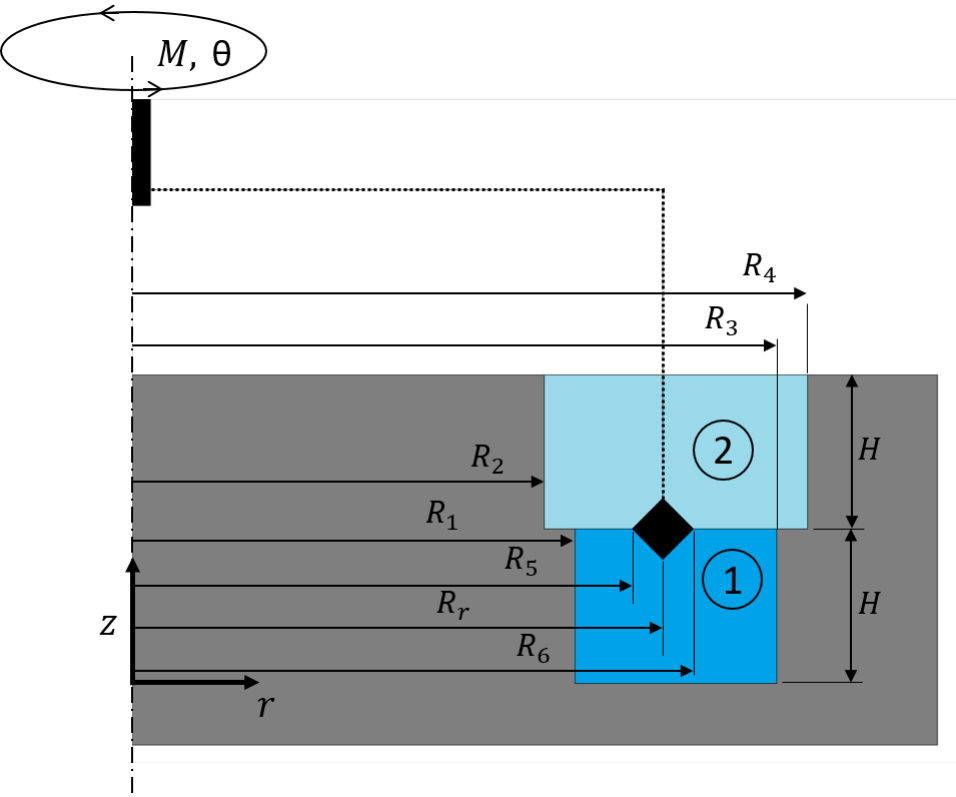}
    \caption{Sketch of the DWR cross-sectional geometry. Only the right half is shown, taking advantage of the rotational symmetry to highlight the details. The different radii are labelled as in reference \protect\cite{Sanchez-Puga2024}.}
    \label{fig:DWRsketch}
\end{figure}

{\bf DWR interfacial shear rheometer.} The DWR geometry comprises two main components \cite{Vandebril2010}: the double wall annular cell, placed in the Langmuir trough, and the ring probe fixture for the commercial Anton Paar rotational rheometers, available at the ILL. The double wall annular cup (see Figure \ref{fig:DWRsketch}), custom made in PTFE, has a double-step radial profile (inner with radius, $R_i = 20$ mm and outer with radius, $R_o = 28.79$  mm) to minimise meniscus effects and ensure interface pinning at the edges of the steps. The double wall annular cell is positioned at the trough's back-end and has two openings, orientated transversely to the barrier motion direction, designed to facilitate a smooth and symmetric entry of interfacial flow into the annular double wall channel. The symmetry axis of the double wall channel is carefully aligned with the rheometer's probe rotation axis. The geometric parameters of the double wall and the ring have been selected in order to: i) make the values of the interfacial shear strain at the inner and outer contact lines at the ring surface as close to each other as possible, and ii) make the distances between the ring and the walls ($3$ mm or more) slightly larger than the air/water capillary length ($\sim 2.7$ mm).

The ring probe has been manufactured using titanium 3D printing technology (3D Systems, Leuven, Belgium), adopting a diamond-shaped cross section \cite{Vandebril2010, Hermans2014} of $1$ mm diagonal. The ring probe is not a closed circle but has three small openings, equally spaced in the angular coordinate, to allow for the inner and outer interfacial regions to be at the same interfacial pressure. The ring probe fixture incorporates a specialised top connection that enables seamless, backlash-free integration with disposable system shafts, ensuring compatibility with Anton Paar rheometers available at the ILL. Accurate centring of the ring probe in the double-wall annular cell is facilitated by the circular shape of the top part of the inner-wall section of the shear cell.

{\bf Integration on FIGARO.}  The considerable length of the trough and the specific horizontal and vertical positioning requirements,  necessary for it to be properly accommodated on the neutron instrument's anti-vibration table, demanded the building of an auxiliary support table. This support table has three legs. The leg placed further from the rheometer rests on an additional plate designed to extend the support surface. This plate effectively enlarges the support area provided by FIGARO's anti-vibration table, ensuring stable and level placement of the trough during measurements.

The neutron beam incidence area has a footprint at the interface $40-60$ mm wide and $80$ mm long in longitudinal and transversal directions to the Langmuir trough, respectively (see Figure \ref{fig:setup_panel}). Then the Langmuir trough is placed so that there is a $10$ mm gap between the end of the annular shear channel ensemble and the beam footprint, in order to minimise possible meniscus effects. Similar $10$ mm gaps were allowed between the neutron beam footprint and the Wilhelmy plate and the mobile barrier at the maximum compression position. 

Finally, to ensure optimal control of the experimental conditions, a cabin has been purposely constructed. The cabin is equipped with lateral quartz windows to facilitate the entry and exit of incident and reflected beams. In addition, the upper wall of the cabin supports a horizontal optical glass window. A laser beam enters the cabin through that window and is used for precise measurement of the vertical distance between the interface and a reference element. This cabin+laser positioning system allows for precise control of the vertical positioning of the interface under study, prevents sample contamination, and enhances the control over the ambient thermodynamic conditions, basically air temperature and relative humidity.

\subsection{Electrical connections and data acquisition}
\label{sec:ElectCon}

\begin{figure}
    \centering
    \includegraphics[width = .7\textwidth]{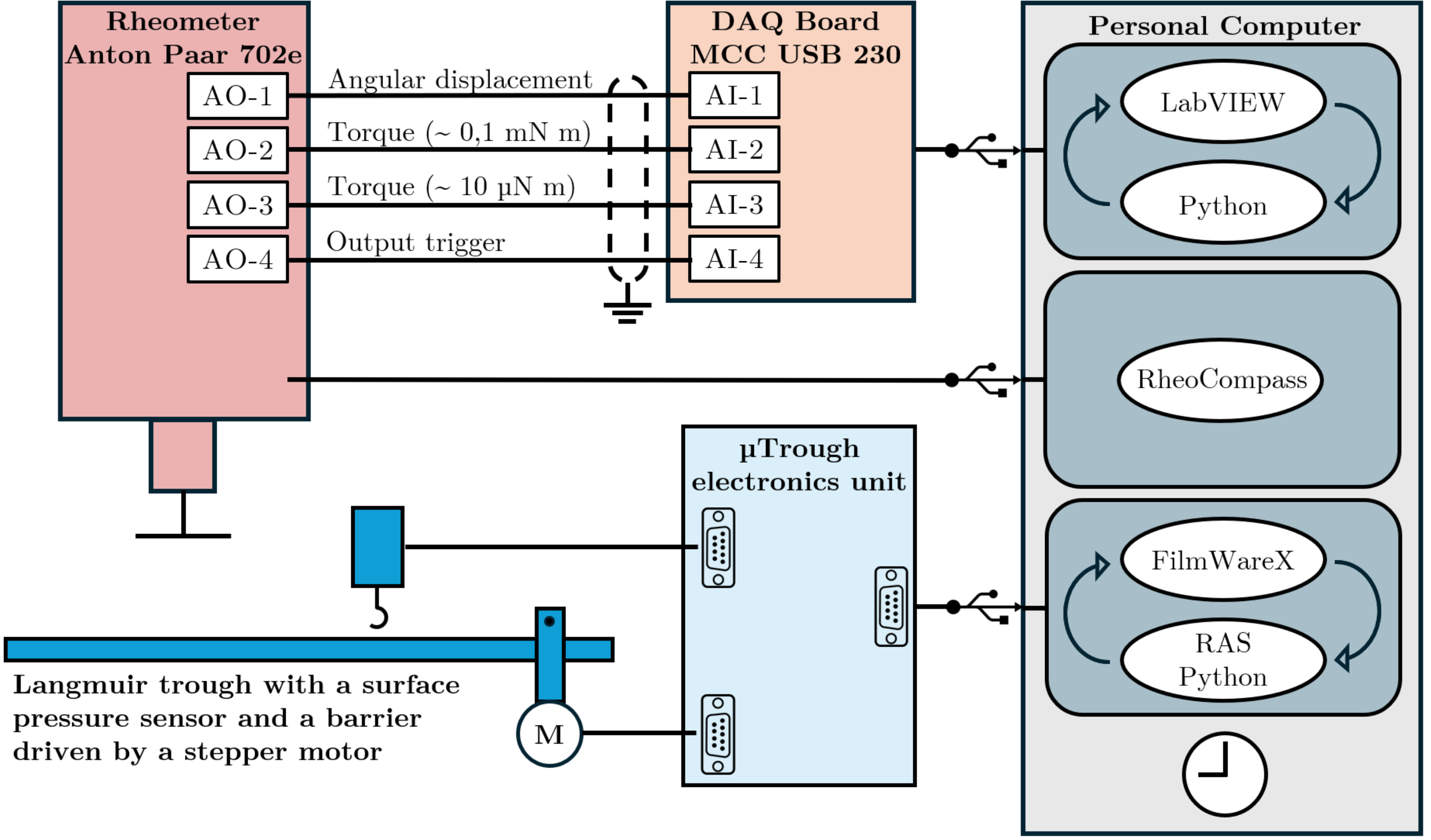}
    \caption{Sketch of the electrical connections. The DAQ board, used to acquire the raw analogue signals from the rheometer, and the control units of the rheometer and Langmuir trough are all connected via USB to a PC running custom-made acquisition/analysis software, rheometer control software, and Langmuir trough control software. The interfacial pressure sensor and the barrier motor of the Langmuir trough are connected to their respective control units via serial connections. The four analogue signals from the rheometer are connected to the DAQ board via the analogue input channels using single-ended connections, with all channels sharing the same ground reference.}
    \label{fig:EleCon}
\end{figure}

In this section, we provide an overview of the electrical connection scheme. Figure \ref{fig:EleCon} illustrates the connections of the interfacial rheology measurement system and the Langmuir trough, both controlled by the same computer.

The Anton Paar MCR702 torsion rheometer here used offers the possibility to configure four analogue output signals ($\pm10$ V and $16$ bit resolution) with selectable gain values according to the user's requirements. In the setup here described, four signals are acquired simultaneously: the angular displacement (gain = 600 V/rad), two torque signals with different gains ($5 \times 10^4$ V/N$\cdot$m and $5 \times 10^5$ V/N$\cdot$m, respectively), and a trigger signal indicating the start of each measurement interval. The two torque signals correspond to the same transducer but are read through separate amplification channels to allow accurate measurements across different orders of magnitude. This trigger signal is used to detect the beginning of a new waveform corresponding to a different measurement to be analysed and stored as raw data for security. These four analogue signals are acquired through a USB DAQ board (Digilent MCC USB-234; 8 SE/4 DIFF analogue inputs; 16-bit resolution, 100 kS/s maximum sampling frequency), which communicates with the control PC via USB. The Anton Paar rheometer is connected via an USB interface to the control tabletop PC where the RheoCompass\textsuperscript{TM} software is run, which controls all the functionalities of the rheometer. The acquisition and analysis of raw signals are performed using custom software developed in LabVIEW\textsuperscript{TM}, which integrates Python subroutines for more complex calculations. 

The electromechanics of the Langmuir trough and the interfacial pressure sensor are connected to a Kibron\textsuperscript{TM} $\mu$Trough measurement and control unit that communicates with the control PC by USB connection. The Kibron proprietary FilmwareX software is mounted on the control PC where it is used to operate the trough. In this case, a Remote Access Server (RAS) has been configured that runs along with FilmwareX\textsuperscript{TM}, sharing access to the trough, allowing communication and operation of the trough using string-based commands and responses. This centralised single-PC control allows us to configure our own scripts in Python to automate experiments and/or perform more complex barrier movement profiles at will.

In this work, the synchronisation of the rheometry and NR measurements was achieved by ensuring that the time clocks of the rheology and NR systems coincide. In the future, it is planned to integrate the Langmuir trough control within NOMAD, a software package developed at the ILL, which allows instrument control and data acquisition. This will make it possible to trigger rheology measurements from NOMAD or, vice versa, to trigger NR measurements from the rheometer control software, in case a very precise synchronisation is needed. For the experiments performed so far, this level of synchronisation has not been necessary.

\subsection{Operation}
\label{sec:operation}

To operate this experimental setup, four distinct software tools are utilised i) to control the rheometer, ii) to acquire the raw torque and angular displacement signals, iii) to manage the Langmuir trough, and iv) to operate the neutron reflectometer. In the following, we will describe some peculiarities of each of these functions.

{\bf Rheometer control.} The correct positioning of the lateral vertices of the ring cross section at the interface level is crucial in this setup. This task can be split into two parts: i) properly defining the vertical length of the probe fixture and the vertical position of the double wall annular shear channel bottom, which is achieved using the rheometer's control software (RheoCompass\textsuperscript{TM}) capability of creating user-defined measurement ensembles, and ii) preparing a vertical positioning script that, starting with the ring above the interface, slowly lowers the probe until a jump in the vertical force measurement is detected. The probe is then lowered another half a millimetre so that the interface pins on the edge of the diamond-shaped cross-section of the ring. Finally, the vertical force measurement is reset. 

It is necessary to properly configure the rheometer control software (RheoCompass\textsuperscript{TM}) for the intended measurements. The user can define different types of test to conduct oscillatory measurements in a single frequency mode, frequency sweep mode, or amplitude sweep mode. The raw torque, angular displacement, and trigger signal are acquired and split into separate waveforms using the trigger signal. 

{\bf Digitising raw rheometry signals.} A software package has been programmed in LabVIEW\textsuperscript{TM} that performs data acquisition and splitting of the acquired signal into individual waveforms. For each individual waveform, an integer number of periods is selected, discarding the initial part, which may contain transients. The waveforms are then processed by discrete Fourier transforms, to obtain the amplitude and phase of both the torque ($T_0$, $\varphi_T$) and the angular displacement ($\phi_0$, $\varphi_\phi$). From there, the complex amplitude ratio follows
 \begin{align}
   AR^* = \frac{T_0\,e^{i(\omega t + \varphi_T)}}{\phi_0\,e^{i(\omega t + \varphi_\phi)}} = \frac{T_0}{\phi_0}e^{i(\varphi_T-\varphi_\phi)} = \left\vert AR^* \right\vert\,e^{i(\varphi_T-\varphi_\phi)},
\end{align}
is calculated, which serves as input for a purpose-built Python programme that implements the corresponding FFBDA scheme and is called from LabVIEW, which yields the interfacial dynamic shear moduli by solving the equations that govern the velocity field in both the interface and the bulk. These analysis tasks are performed asynchronously in parallel using a queue system as acquisitions are made.

{\bf Langmuir trough management.} Kibron\textsuperscript{TM} components were used in the assembly of the Langmuir trough. Consequently, the Langmuir trough is operated using the company's proprietary software, FilmwareX. This software includes a specialised feature that allows the integration of Python scripts, enabling users to employ various operational modes through the Remote Access Server on a local network. This capability offers significant flexibility, allowing users to automate the measurement process in coordination with neutron scattering data acquisition. In addition, it facilitates the development of intricate barrier movement profiles, thereby enhancing the precision and complexity of experimental setups.

{\bf Operation of the neutron reflectometer.}The FIGARO instrument at the Institut Laue-Langevin is a high-flux time-of-flight (TOF) reflectometer \cite{Campbell2011}. It is equipped with four choppers that allow one to select the wavelength resolution. In this study, for example, it was used with a constant resolution d$\lambda$/$\lambda = 7$ \%. Three different incidence angles $\theta_1$ = 0.62 $^\circ$, $\theta_2$ = 1.97 $^\circ$ and $\theta_3$ = 3.8 $^\circ$, can be configured. Typically, when seeking a measurement that spans a broad wavelength range, it is preferable to measure at angles $\theta_1$ and $\theta_3$, because their ranges overlap. However, for rapid measurements in studies on kinetic processes \cite{Campbell2018b}, measurements can be made at a single incidence angle tailored to the user's momentum transfer range requirements. The instrument also features the ability to measure surface excess with high precision by measuring at low Q$_z$ ($\theta_1$) on a mixture of heavy and light water with zero neutron reflection \cite{Braun2017,Campbell2018b}. This makes it possible to quantify the composition of binary mixtures using isotopic contrasts in a suitable manner or in combination with ellipsometry as a complementary technique. On the other hand, when the objective of the study is to obtain information on the evolution of the interfacial structure, it is possible to make measurements using higher incidence angles ($\theta_2$ or $\theta_3$) depending on the range in Q$_z$ where the characteristics of the changes in the reflectivity curve appear \cite{Carrascosa-Tejedor2022}. Different isotopic contrasts can be tailored to match the structural complexity of the system under study, and, using deuteration, distinct components can be selectively highlighted. The two-dimensional detector signal is processed and reduced using COSMOS \cite{Gutfreund2018}, ultimately yielding neutron reflectivity $R$, with respect to the vertical scattering vector $Q_z$ which, considering specular reflection, reads as follows:
\begin{align}
    Q_z = \frac{4\pi}{\lambda}\sin{\theta},
\end{align}
where $\theta$ is the angle of incidence and $\lambda$ is the wavelength. Specular neutron reflectometry provides detailed information on layered structures perpendicular to the interface. The experimental data are typically analysed using an optical matrix formalism, in which the reflectivity is calculated for a stack of layers, each characterised by a specific scattering length density (SLD), thickness, roughness, and solvent volume fraction. To ensure physical consistency, constraints can be applied within the modelling software. By fitting datasets acquired under different isotopic contrast conditions using a shared structural model, one can determine both the composition and depth profile of the interfacial material with sub-nanometric precision.

\subsection{Data analysis}

{\bf Rheology data analysis.} In fluid-fluid interfaces, the contributions of interfacial and bulk phases are intrinsically coupled. Hence, the only way to correctly decouple both effects is to work with the flow fields at the interface and the bulk fluid phases. Obviously, this makes the task quite complicated compared to analytic calculations. However, several FFBDA schemes have recently been proposed for different ISR configurations, either in longitudinal \cite{Reynaert2008,Tajuelo2015,Tajuelo2016} or rotational motion \cite{Vandebril2010,Tajuelo2018,Sanchez2021}. In all of them, a simple physical model of the flow geometry allows the user to formulate the Navier-Stokes equations for bulk fluid flows with just one velocity component (in rotational rheometers, such as the DWR, the azimuthal one) \cite{Vandebril2010,Sanchez-Puga2024, DWR_psp_mar}. Then, the stress balance at the interface is included through the Boussinesq-Scriven equation \cite{Reynaert2008,Vandebril2010}, considering that only shear stresses occur at the interface and the bulk fluids. The crucial parameter in this problem appears in the Boussinesq-Scriven equation, namely the complex Boussinesq number, $Bq^*$, which describes the relative importance of the interfacial drag compared to the bulk drag. For the most common case of Newtonian bulk fluid phases and linear viscoelastic interfaces, $Bq^*$ is defined as \cite{Edwards1991}:
\begin{align}
    Bq^* = \frac{\eta_s^*\frac{V}{L_s}P_s}{\eta\frac{V}{L_b}A_b} \simeq \frac{\eta_s^*}{\eta a},
    \label{eq:Boussi}
\end{align}
where $\eta$ and $\eta_s^*$ are the subphase and complex interfacial viscosities, $V$ is the characteristic velocity, $L_s$ and $L_b$ are characteristic length scales for the decay of linear momentum at the interface and in the bulk fluid, respectively, $P_s$ and $A_b$ are the perimeter of the contact line at the probe surface and the area of contact between the probe and the bulk subphase, and $a$ is a length scale defined by the probe's area-to-perimeter ratio, $a = A_b/P_s$. In the present case, a ring with diamond cross-section with side $L$, it is usual to consider $a = L$ (for the DWR on the air/water interfaces $L \sim 0.7$ mm \cite{Vandebril2010, Renggli2020, Sanchez-Puga2024}). Strictly speaking, $L_b$ and $L_s$ are frequency-dependent scale lengths \cite{Fitzgibbon2014}, so that $Bq^*(\omega) \sim L_b (\omega)/L_s (\omega) \simeq \omega^{-1/4}$. As the typical frequency range in interfacial rheology measurements is not very wide, this dependence can be safely ignored.

When $Bq^*$ is large (say $Bq^* \ge 100$) interfacial stresses dominate and simple expressions \cite{Sanchez2021} can be used to obtain the value of $Bq^*$ from the experimental data and, consequently, of $\eta^*_s$. Unfortunately, this is not always the case, and then it is necessary to properly analyse the data to separate the contributions of the strongly coupled interfacial and bulk flows. 

In that purpose, an iterative procedure is established that involves: 1) solving the Navier-Stokes equations together with the Boussinesq-Scriven equation starting from an initial value \emph{seed} of $Bq^*$, 2) obtaining the values of the interfacial and bulk drags for that flow configuration, and 3) using the drag values and the experimental value of the complex amplitude ratio, $AR^*$, to obtain a corrected value for $Bq^*$ through an iterative scheme.

The equation of motion of the DWR can be written as
\begin{align}
    T_0\,e^{i(\omega t + \varphi_T)} + T_s(t) + T_1(t) + T_2(t) = I\ddot{\phi}(t),
    \label{eq:eq_motion}
\end{align}
where the first term is the torque imposed by the motor on the DWR assembly, the next three terms are the torques due to the drag from the interface, bulk phase 1 and bulk phase 2, respectively, and $I$ is the moment of inertia of the rotor and the DWR assembly. Assuming all terms in Eq. \eqref{eq:eq_motion} are oscillatory with frequency $\omega$, the time dependence vanishes out and we can rewrite the equation of motion as
\begin{align}
   T_0^* + T_s^* + T_1^* + T_2^* = -\,\omega^2\,I\,\phi_0^*,
\end{align}
where the argument of each (complex) term accounts for the phase difference with respect to a given reference. Both $T_0^*$ and $\phi_0^*$ are measured during the experiment. $T_s^*$ is proportional to $Bq^*$, and both $T_1^*$ and $T_2^*$ depend on the flow field, which in turn depends on $Bq^*$. Therefore,  $Bq^*$ cannot be directly calculated from the equation of motion and an iterative scheme along the lines previously explained is necessary.

\cite{Vandebril2010} first published and made freely available an FFBDA software package (\url{https://softmat.mat.ethz.ch/opensource.html}) specifically written for DWR interfacial shear rheometer configurations. In the present case, we used to analyse the data a second generation FFBDA software package (freely available too at \url{https://github.com/psanchez0046/DWR-Drag} and \url{https://doi.org/10.5281/zenodo.16459609}) that incorporates the following improvements: i) a user-selectable increased mesh resolution, ii) a second-order finite difference approximation for drag calculations, and iii) an iterative scheme based on the probe's equation of motion. Full details can be found in \cite{Sanchez-Puga2024}.

It is important to realise that in the DWR configuration the interfacial strain field can be considered uniform only for high values of $Bq^*$ (say $Bq^* \ge 100$). For lower values of $Bq^*$, the interfacial strain field is highly non-linear, with the highest strain values just at the probe interface contact line. After convergence, the FFBDA software package can yield the strain value on the contact line. However, for labelling purposes, in the rest of this report, we will use an average strain value, $\gamma_{av}$, obtained from the analytic solution of the Boussinesq-Scriven equation in the limit $Bq^* \xrightarrow{}$ $\infty$, which yields
\begin{align}
    \gamma_{av} = \phi_0 \frac{R_3^2}{R_3^2 - R_{6}^2} \sim \phi_0 \frac{R_1^2}{R_{5}^2 - R_{1}^2}
\end{align}
where $\phi_0$ is the amplitude of the angular displacement oscillation and the different radii are as indicated in Figure \ref{fig:DWRsketch}.

{\bf NR data.} The measured reflectivity curves (Fig. \ref{fig:NR}) have been fitted using \textsf{refnx} software \cite{Nelson2019}, which allows modelling of the interfacial structure in terms of several piled-up layers and contains a versatile environment for model refinement and implementation of constraints. For instance, in the case of lipid monolayers, the interface model can be established straightforwardly using a specific macro for the modelling of lipid layers (the `LipidLeaflet' macro). This macro implements the molecular constraint that ensures the same area per molecule ($A_{\textrm{molec}}$) in both the headgroup layer and the tail layer, in which the lipid monolayer is sub-divided \cite{Gerelli2016, Campbell2018a, Nelson2019}. 
After the fitting procedure, one can obtain a scattering length density (SLD) profile across the vertical coordinate and the values of the corresponding parameters defining the transverse structure of the interfacial film (see Table \ref{tab:summ}).

\section{Materials and experimental protocol}
% \label{}

The performance of the setup was assessed by measuring Langmuir DPPC monolayers at different interfacial pressures at $22$ $^\circ$C. Chain-deuterated DPPC (d$_{62}$-DPPC) was received from Avanti Polar Lipids ($>$ $99$ \%). Subphase water, H$_2$O, was obtained through a Milli-Q dispenser (Millipore) and D$_2$O was used as received from Sigma-Aldrich. The Langmuir trough, the annular shear channel, and the lateral barrier were meticulously cleaned with chloroform (Sigma Aldrich). All of these components were carefully rinsed with water to remove any remaining residues. The DWR probe was submerged in chloroform about $5$ min before each experiment for cleaning.

The chloroform solutions of the lipids were prepared at concentration $0.5$ mg/mL and gently spread drop-wise on the interface using a Hamilton micro-syringe until the interfacial pressure is approximately $2$ -- $3$ mN/m. The chloroform is then allowed to evaporate and the monolayer is left to equilibrate for $15$ min. Then sequential increases in interfacial pressure, using the constant pressure mode of operation between steps, allows at each step the acquisition of full $Q_z$ range NR measurements simultaneously with a set of continuous rheological measurements at a single frequency ($3$ \% strain and $0.5$ Hz), followed by measurements during a frequency sweep ($0.3$ -- $3$ Hz at $3$ \% strain) and a strain sweep ($1$ -- $10$ \% at frequency $0.5$ Hz). 

Finally, all the measurements shown in this paper have been performed in strain-controlled mode (TruStrain\textsuperscript{TM}) to avoid excessive strains that could extremely shear the sample. However, we point out that in the rotational rheometer here used this strain control mode is implemented through a rather fast feedback loop that governs the electromechanical torque that is imposed on the probe+rotor ensemble.

% \section{Performance}
% \label{sec:performance}

\section{Experimental validation of instrument performance}
\label{sec:performance}

Simultaneous neutron reflectometry and interfacial shear rheology measurements were carried out on DPPC monolayers at constant interfacial pressure to test the performance of the experimental setup. DPPC is one of the most studied phospholipids, so there exists an extensive literature using neutron reflectivity \cite{Campbell2018a, Carrascosa2020} and interfacial shear rheology \cite{Kim2011, Hermans2014} that allows us to compare our measurements. Deuterated DPPC was used, with Air Contrast Matched Water (ACMW, a mixture of  8.2$\%$ D$_2$O and 91.8$\%$ H$_2$O leading to a SLD = 0 that matches the air layer) and pure subphases of D$_2$O subphases. Measurements have been made at relatively high interfacial pressures ($25$, $35$ and $45$ mN/m) away from the liquid expanded to liquid condensed (LE-LC) phase transition (about $8$ mN/m) \cite{Campbell2018a}, but still yielding $G_s^{\prime\prime}(\omega)$ values low enough to show the instrument's resolution limit.

\begin{figure}
    \centering  
    \includegraphics[width=\linewidth]{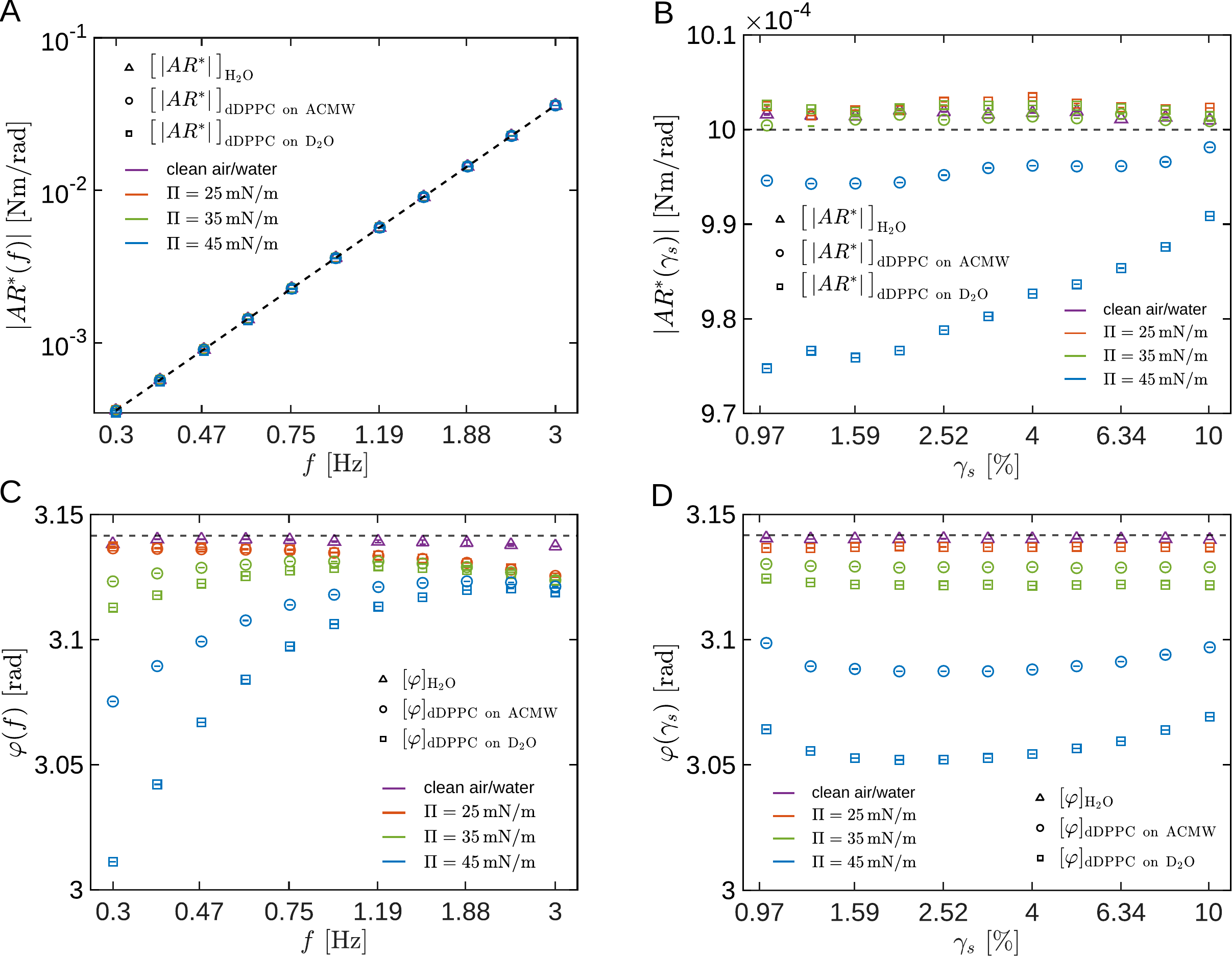}
    \caption{Panel A: $|AR^*(f)|$ at $\gamma = 3$ \%. Panel B: $|AR^*(\gamma_s)|$ at $f = 0.5$ Hz. Panel C: $\varphi(f)$ at $\gamma = 3$ \%. Panel D: $\varphi(\gamma_s)$ at $f = 0.5$ Hz. DPPC monolayers on ACMW (circles) and D$_2$O (squares) subphases, at $\Pi = 25$ mN/m (red), $\Pi = 35$ mN/m (green), and $\Pi = 45$ mN/m (blue). Clean air/water interface is represented with purple triangles. A dashed line with slope $2$ in panel A, a dashed horizontal line at $10^{-3}$ in panel C and a dashed horizontal line at $\pi$ in panels B and D have been plotted to guide the eye.}
    \label{fig:AR}
\end{figure}

{\bf Validation of the DWR ISR.} In panels A and C of Figure \ref{fig:AR} we show the frequency dependence (where $f$ is the frequency in Hz) of the modulus of the complex amplitude ratio, $|AR^*(f)|$ (panel A) and the phase lag between the torque and the angular displacement signals, $\varphi(f)$ (panel C), on the oscillation frequency, at the same interfacial strain $\gamma_s = 3$ \%. The data belong to a clean air/water interface and DPPC monolayers on two subphases, with different NR contrasts (ACMW and D$_2$O, respectively), at three different interfacial pressures: $\Pi = 25$ (red), $35$ (green) and $45$ (blue) mN/m. The uncertainty in all the interfacial rheology data shown here has been estimated as indicated in Section 1 of the Supporting Information.

The values of the modulus of the amplitude ratio, $|AR^*(f)|$, show similar trends for all the investigated interfaces and are practically indistinguishable in the logarithmic representation. Moreover, the values of the modulus of the amplitude ratio show a clear trend $f^{2}$ that indicates that the system is working in a regime where the effects due to the inertia of the rotor+probe ensemble are relevant. Furthermore, the phase lag graphs between the applied torque and the angular displacement, $\varphi(f)$, show in all cases smooth low-noise curves with values close to $\pi$ rad, as expected when the rotor+probe inertia is important. For this geometry and the frequency range explored here, the two most important contributions to $AR^*$ are the interfacial effects and the inertia of the instrument. Interestingly, the $\varphi(f)$ curves approach each other at high frequency and apparently show a decreasing trend above a frequency that depends on the specific interfacial system (surfactant and bulk fluid phases). In any case, data for $f \gtrsim 1$ Hz should be considered with some reserve.

In panels B and D of Figure \ref{fig:AR} we show the dependence of the modulus of the complex amplitude ratio, $|AR^*|$ (panel B) and the phase delay between the torque and angular displacement signals, $\varphi$ (panel D), on the interfacial strain, $\gamma_s$, at a constant frequency, $f = 0.5$ Hz. The data correspond to a clean air/water interface and DPPC monolayers on two subphases, with different NR contrasts (ACMW and D$_2$O, respectively), at three different interfacial pressures: $\Pi = 25$ (red), $35$ (green) and $45$ (blue) mN/m. The limited amount of time allocated at the ILL for these experiments did not allow measurement of the strain sweep in the DPPC on ACMW subphase at $\Pi = 25$ mN/m. Note that the error bars are smaller than the symbols.

The measured values for $|AR^*|$ or $\varphi$ show a very small dependence on the strain. Notice that, in contrast to Figure \ref{fig:AR}A, here the vertical axis scale is linear, and all the values represented here differ at most in less than $4$ \%. The data in Fig. \ref{fig:AR}B corresponding to the interfaces with a stronger rheological response (DPPC at $45$ mN/m) show smaller $|AR^*|$ than those corresponding to a clean interface, which can seem counter-intuitive considering that $AR^*$ is defined as $T_0^*/\phi_0^*$ (the more rheologically responsive DPPC at $\Pi = 45$ mN/m interface seems to require a lower torque to be sheared than that corresponding to a clean interface). However, a viscoelastic interface in an oscillatory experiment can give rise to a resonant response. In the high $Bq^*$ regime, Eq. \eqref{eq:eq_motion} can be approximated as
\begin{align}
    T_0e^{i(\omega t + \varphi_T)} + T_s(t) = I\ddot{\phi}(t)\label{eq:motion_approx}
\end{align}
where the torque resulting from the drag from the interface is proportional to $-\eta_s^*\dot{\phi}(t)$
\begin{align*}
    T_s(t) = -i\,\omega\,C_g\,\eta_s^*\phi_0\,e^{i(\omega t + \varphi_\phi)},
\end{align*}
being $C_g$ a positive geometric coefficient \cite{Renggli2020,Sanchez2021}. From Eq. \eqref{eq:motion_approx} and considering $\eta_s^*=\eta_s^\prime-i\eta_s^{\prime\prime}$, the amplitude ratio is given by
\begin{align}
    AR^*= i\,\omega\,\eta_s^\prime\,C_g + \omega\,\eta_s^{\prime\prime}\,C_g - I\,\omega^2,
\end{align}
so that its modulus is
\begin{align}
    \vert AR^* \vert = \sqrt{(\omega\,\eta_s^{\prime\prime}\,C_g - I\,\omega^2)^2 + (\omega\,\eta_s^\prime\,C_g)^2}.
    \label{eq:AR_modulus_approx}
\end{align}

From Eq. \eqref{eq:AR_modulus_approx} it is now apparent that, in a system where the governing contributions are the inertia and the interfacial drag, $\vert AR^* \vert$ shows a minimum when the interfacial storage modulus is such that $\omega\,\eta_s^{\prime\prime}\,C_g - I\,\omega^2=0$ (a detailed description of the corresponding second order dynamic model is provided in Section 2 of Supporting Information). Therefore, the raw data shown in Fig. \ref{fig:AR}B suggest, prior to any data analysis, that the DPPC interfaces at $45$ mN/m must have a measurable storage modulus and, furthermore, that its value must change with strain amplitude.

\begin{figure}
  \centering  
  \includegraphics[width=\linewidth]{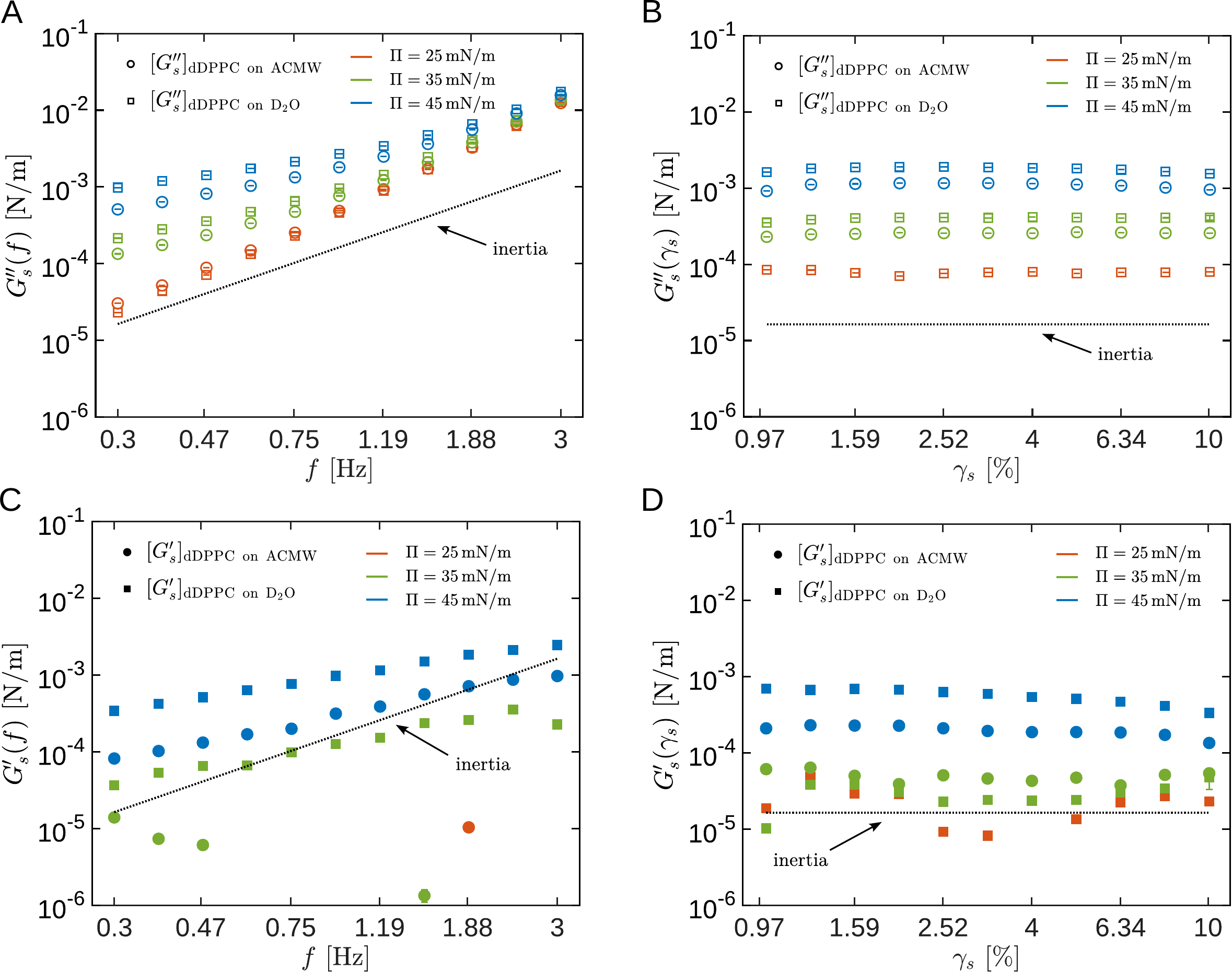}
  \caption{Panel A: Loss modulus, $G_s^{\prime\prime}(f)$ at $\gamma_s = 3$ \%. Panel B: Loss modulus, $G_s^{\prime\prime}(\gamma_s)$ at $f = 0.5$ Hz. Panel C: Storage modulus, $G_s^{\prime}(f)$ at $\gamma_s = 3$ \%. Panel D: Storage modulus, $G_s^{\prime}(\gamma_s)$ at $f = 0.5$ Hz. DPPC monolayers onto ACMW (circles) and D$_2$O (squares) subphases, at $\Pi = 25$ mN/m (red), $\Pi = 35$ mN/m (green), and $\Pi = 45$ mN/m (blue). The dotted line indicates the inertia-limited sensitivity.}
  \label{fig:Gs}
\end{figure}

The loss and storage modulus corresponding to the measurements shown in Figure \ref{fig:AR} are shown in Figure \ref{fig:Gs}, where the dotted lines labelled \textit{inertia} indicate that below that line the inertia of the rotor+probe dominates and limits the operational window of the instrument \cite{Renggli2020}. In other words, the dotted lines represent the conditions in which the dynamic moduli values ($G^\prime_s$ or $G^{\prime\prime}_s$) are equal to
\begin{align}
    \frac{I\omega^2}{C_\Phi/C_M},
\end{align}
where $C_\Phi/C_M$ is a geometric coefficient \cite{Renggli2020,Sanchez2021}, that is,
\begin{align} 
  \frac{C_\Phi}{C_M} = 4\pi\left(\frac{R_5^2 R_1^2}{R_5^2 - R_1^2} + \frac{R_6^2 R_3^2}{R_3^2 - R_6^2}\right).
  \label{eq:C2}
\end{align}

In the plots of the storage modulus, Figures \ref{fig:Gs}C adn D, the inertia lines are too close to most of the $G^\prime_s$ data, but the ones at $\Pi = 45$ mN/m. Hence, probably only the values of the storage modulus measured here at $\Pi = 45$ mN/m are reliable.

Regarding the frequency dependence in Figure \ref{fig:Gs}A and C, typically, the calculated values of the storage modulus, $G_s^\prime(f)$, are smaller than those of the loss modulus, $G_s^{\prime\prime}(f)$, under the same conditions of subphase, interfacial pressure, frequency, and strain. In Figure \ref{fig:Gs}A most data, but perhaps those at $\Pi = 25$ mN/M and the lower frequencies, are far enough from the inertia limit. However, in Figure \ref{fig:Gs}C only the $G_s^{\prime}(f)$ data corresponding to $\Pi = 45$ mN/m appear to be far enough from the inertia limit to be considered reliable. It is worth noting that at 45 mN/m the elastic component, $G_s^{\prime}(f)$ is smaller than the viscous one, $G_s^{\prime\prime}(f)$, but becomes significant, approaching the same order of magnitude as the viscous component.

Several other aspects of the frequency dependence of the loss modulus (see Figure \ref{fig:Gs}A) can be observed. First, the loss modulus increases with the interfacial pressure, as expected. Second, the monolayers onto a D$_2$O subphase show loss modulus values systematically higher than the monolayers loaded on ACMW subphases. Third, all curves converge at high frequency, as expected from the curves shown in panels A and C of Figure \ref{fig:AR}.

Regarding the dependence of strain amplitude (at $f = 0.5$ Hz), shown in panels B and D of Figure \ref{fig:Gs}, the loss modulus is in all cases above the inertia limit and is fairly constant for all strain amplitude values within the explored range (1-10\%) and any of the interfacial pressures reported. Moreover, panels B and D of Figure Figure \ref{fig:Gs} show that for 35 and 45 mN/m interfacial pressure $G_s^{\prime\prime}(\gamma)$ is consistently higher that $G_s^{\prime}(\gamma)$, confirming the  viscosity dominated character of the monolayer. 
Therefore, the strain amplitude value used in the frequency sweeps ($\gamma_s = 3\%$) can be safely assumed to lie within the material's linear regime with the current experimental configuration and for the DPPC monolayers investigated in this study. Measurements at strains below 1\% are not technically feasible due to the inertia of the motor drive, which represents a limiting factor in interfacial rheometry. The storage modulus for the weaker interfaces ($\Pi = 25$ and $35$ mN/m) is closer to the inertia lilmit, which means that the elastic response of these interfaces is close to the instrument sensitivity. For the more responsive interfaces ($\Pi = 45$ mN/m), $G_s^\prime$ is well above the inertia limit and a slight decrease can be observed for $\gamma_s$ above $4$ \%, in good agreement with the previous analysis of the raw $\vert AR^* \vert$ data.

From a physical perspective, at such a high interfacial pressure the monolayer is expected to be in the LC phase. Although the lateral diffusion of phospholipid molecules is significantly reduced compared with the LE phase, the system still appears to retain a fluid character, with some molecular motion and dissipative processes remaining possible. Such lateral mobility should be higher at lower interfacial pressures and, therefore, the storage modulus would be expected to decrease faster than the loss modulus upon decreasing the interfacial pressure. Such a tendency is confirmed by the strain dependence measurements shown in Figures \ref{fig:Gs}B and D.
Hence, it is not surprising that, at the interfacial dynamical conditions used here, the interface exhibits a predominantly viscous behaviour. Such behaviour is also consistent with previous reports in the literature. For instance, \cite{Espinosa2011, Kim2011, Kim2013} have shown that DPPC monolayers in condensed phases exhibit fluid-like behaviour. More recently, \cite{Hermans2014} describe DPPC interfaces as dominated by viscous behaviour, for the same pressure, frequency, and temperature ranges studied here.

{\bf Simultaneous neutron reflectometry measurements.} The data for the two isotropic contrasts were co-refined under the assumption that the chemical structures are identical in both cases. The thickness of the acyl chains is $t_\mathrm{AC} = \frac{V_{\mathrm{AC}}}{A_\mathrm{molec}}$, as we assume full occupancy of the tail group and the thickness of the phosphatidylcholine (PC) headgroup is $t_\mathrm{PC} = \frac{V_{\mathrm{PC}}}{A_\mathrm{molec}\cdot(1 - \phi_\mathrm{PC, w})}$, where $\phi_\mathrm{PC, w}$ is the hydration fraction of the headgroup.

The values of the parameters used in the two-layer model are shown in Table \ref{tab:summ}. The background for the ACMW contrast was fixed at $5\times10^{-6}$, and for the D$_2$O contrast at $10^{-7}$. The molecular volume of the acyl chain was fixed to that corresponding to the LC phase ($V_{\mathrm{AC}} = 759$ $\mathrm{\AA}^3$), the molecular volume of the PC headgroup was fixed to $V_{\mathrm{PC}} = 344$ $\mathrm{\AA}^3$, and the thickness of the headgroup was fixed to $9$ $\mathrm{\AA}$, all taken from reference \cite{Campbell2018a}. The interfacial roughness ($\sigma$) was fixed to the value corresponding to the capillary waves according to reference \cite{Ocko1994}. The modest increase in interfacial roughness from $3.45$ $\mathrm{\AA}$ to $4.53$ $\mathrm{\AA}$ upon compression may signal the onset of out-of-plane fluctuations or molecular protrusions as the monolayer approaches its collapse pressure.

The SLD of D$_2$O, $\mathrm{SLD}_{\mathrm{D_2O}}$, was treated as a free parameter and shows decreasing values due to exchange with atmospheric water. The observed progressive increase in acyl-chain thickness from $16.27$ to $17.41$ $\mathrm{\AA}$ upon compression from $25$ to $45$ mN/m indicates a slight chain extension and vertical orientation. The modest magnitude of this change, despite a $\sim7$\% reduction in area per molecule, suggests that the chains are already well-oriented at $25$ mN/m and that further compression primarily reduces lateral packing defects rather than driving additional chain ordering. The large acquisition times needed for the NR measurements preclude the possibility of making continuous isothermal compressions. However, the fitted molecular areas are in good agreement with previous reference continuous isotherms of hydrogenous and deuterated DPPC monolayers (see, for instance, Figure 5A in \cite{Campbell2018a}). In general, the two-layer model parameters obtained are in good agreement with those previously reported from independent FIGARO measurements \cite{Campbell2018a, Carrascosa2020}.

\begin{figure}
  \centering  
  \includegraphics[width=.5\linewidth]{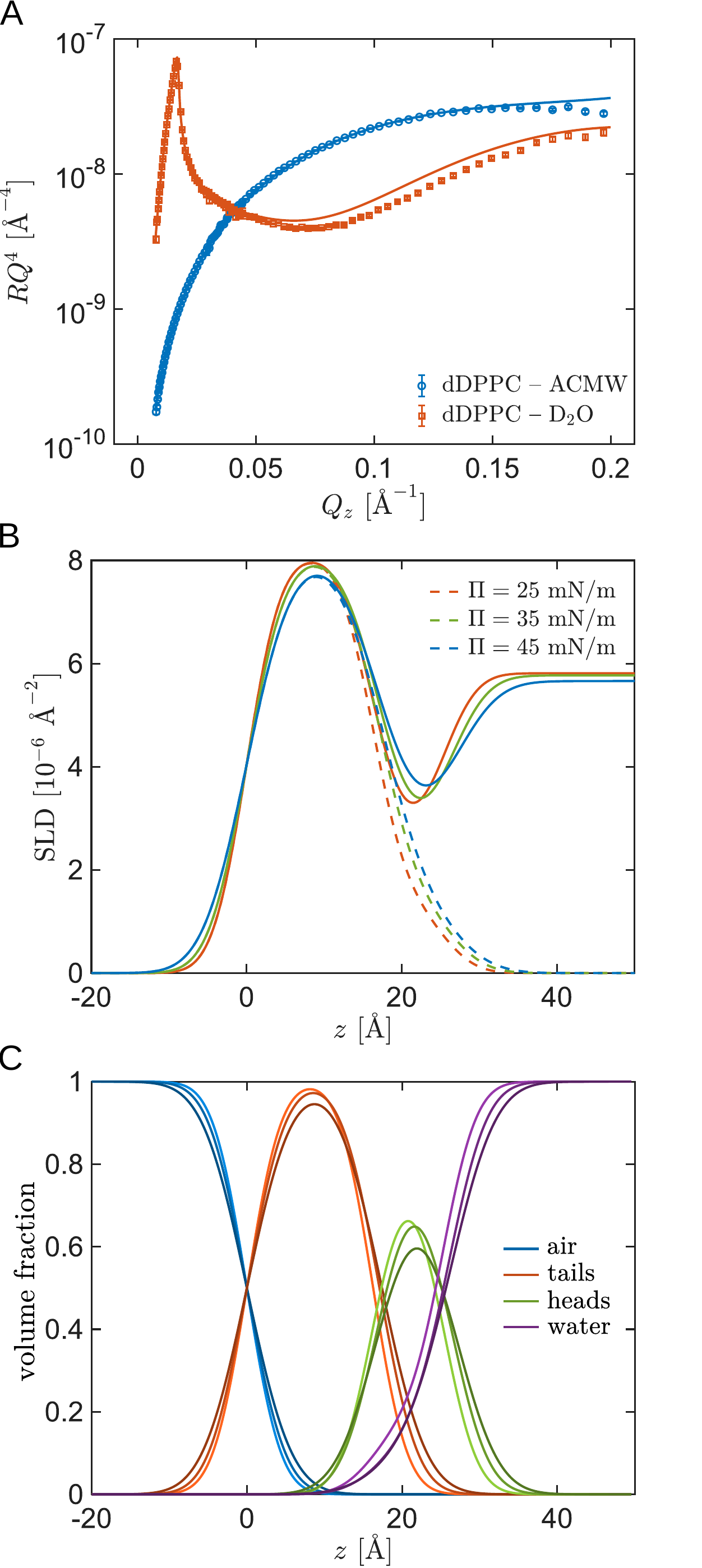}
  \caption{Panel A: $RQ^4$, as a function of the vertical scattering vector, $Q_z$, at $\Pi = 25$ mN/m for the two different subphases used here. Panel B: Calculated SLD profiles at $\Pi = 25$ mN/m (red lines), $\Pi = 35$ mN/m (green lines) and $\Pi = 45$ mN/m (blue lines); dashed lines: ACMW subphase; continuous lines: D$_2$O subphase. Panel C: Volume fraction profiles of different slabs: air (blue), tails (red), heads (green), and water (purple). Lighter traces correspond to lower interfacial pressure.}
\label{fig:NR}
\end{figure}

In Figure \ref{fig:NR} we show illustrative examples of the data obtained through the NR measurements. Figure \ref{fig:NR}A shows an example of the reflectivity curves at $\Pi = 25$ mN/m for the two measured contrasts (see Supporting Information for the corresponding curves at $\Pi = 35$ and $45$ mN/m). Figures \ref{fig:NR}B and \ref{fig:NR}C display, respectively, the calculated SLD profiles with the different contrasts (subphases) at different interfacial pressures and the corresponding volume fraction of the defined slabs in the vertical coordinate at different interfacial pressures. The origin of the vertical coordinate, $z$, is set at the air-tails dividing surface.

The fit of the reflectivity curve corresponding to the interfacial pressure of $25$ mN/m is shown in Figure \ref{fig:NR}A. The corresponding SLD profile and the volume fraction representation (Figures \ref{fig:NR}B and \ref{fig:NR}C) at different interfacial pressures reveal systematic trends with increasing interfacial pressure that can be correlated to the values reported in Table \ref{tab:summ}: i) the peak and valley positions, corresponding to the tail and head regions, respectively, shift upward owing to the slight increase in tail thickness and ii) the SLD in the head region increases due to a decrease in the water volume fraction. This trend reflects changes in the hydration of the lipid headgroup: as the interfacial pressure increases, the water content in the headgroup decreases, leading to a higher SLD. The systematic decrease in headgroup hydration upon compression from 18\% at $25$ mN/m to 12\% at $45$ mN/m water volume fraction reflects progressive dehydration of the phosphatidylcholine moieties as intermolecular spacing decreases. Importantly, the persistence of $\sim12$\% water even at the highest pressure indicates that the headgroups retain a hydration shell that enables molecular rearrangement under shear. This residual hydration is consistent with the predominantly viscous rheological response ($G_s^{\prime\prime}$ $>$ $G_s^\prime$) observed, as complete dehydration would be expected to yield a more elastic, solid-like behaviour characteristic of a true solid phase.

\begin{figure}
  \centering  
  \includegraphics[width=.5\linewidth]{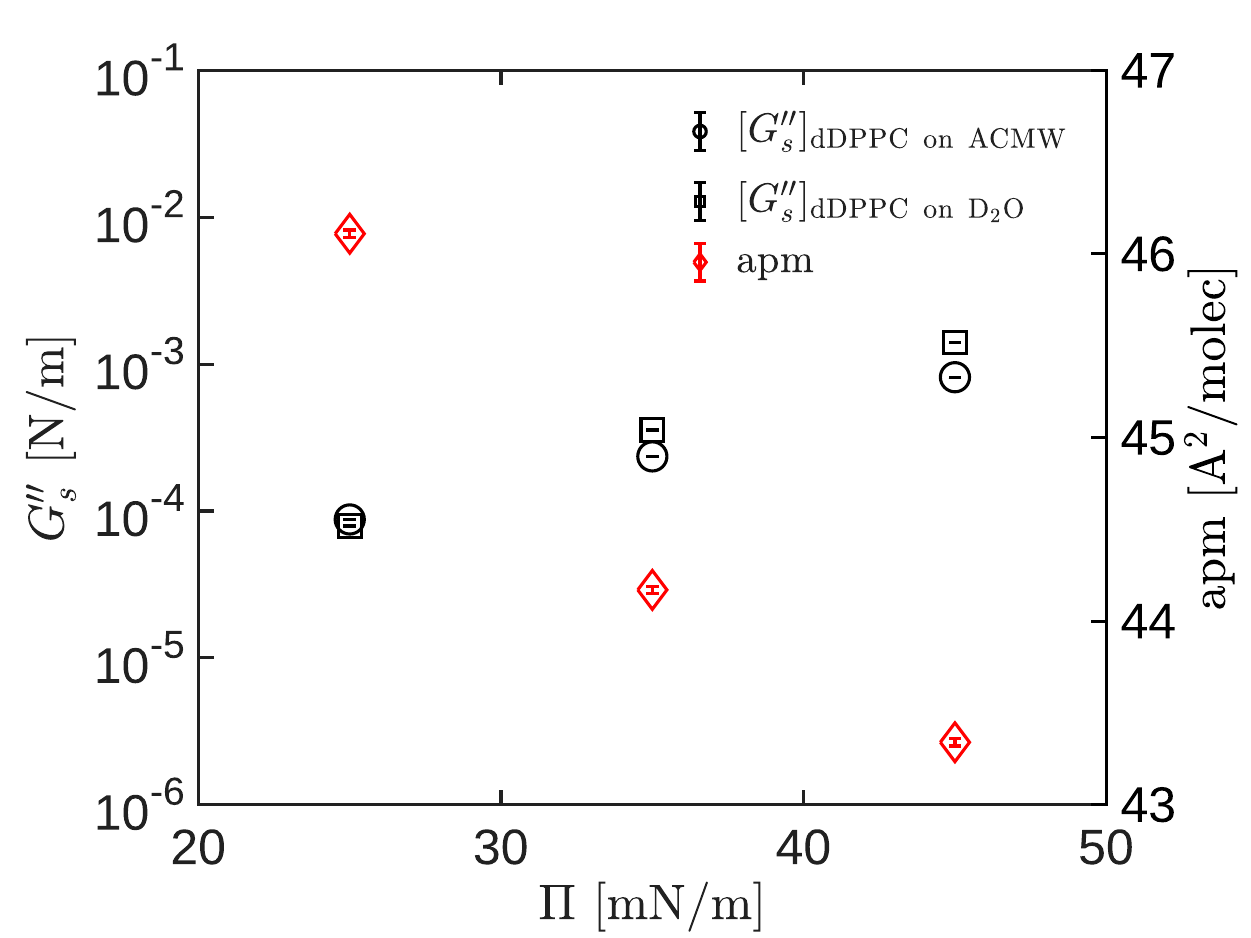}
\caption{Simultaneous measurements of the loss modulus (left axis; black symbols), $G_s''$, and the area per molecule (right axis, red diamonds) at $f = 0.5$ Hz and $\gamma_s = 3$ \%, at three interfacial pressure values. Circles and squares correspond to DPPC monolayers onto ACMW and D$_2$O subphases, respectively.}
\label{fig:summ}
\end{figure}

Finally, in Figure \ref{fig:summ} we put together the information obtained from the simultaneous measurement of interfacial rheology and neutron reflectometry, with ACMW and D$_2$O subphases. In Figure \ref{fig:summ}, we show the graphs of the loss modulus $G_s^{\prime\prime}$ (left axis, black symbols) and the area per molecule $A_{\textrm{molec}}$ (right axis, red symbols), as a function of interfacial pressure, respectively. All rheological measurements were taken at $f = 0.5$ Hz and $\gamma_s = 3$ \%, at three interfacial pressure values. The values shown are the averages of at least $20$ measurements, and the error bars represent the standard error of the mean. The circles and squares refer to the DPPC monolayers on the ACMW and D$_2$O subphases, respectively. For DPPC monolayers, an exponential relationship between the loss modulus and the interfacial pressure can be observed (see Figure \ref{fig:summ}). The measured values of the dynamic moduli agree well with other previous studies \cite{Kim2011, Kim2013, Hermans2014}. Moreover, the dependencies of $G^{\prime\prime}_s$ and $A_{\textrm{molec}}$ on interfacial pressure are consistent with each other: the higher the interfacial pressure, the higher the dynamic moduli, and the lower the mean area available for the surfactant molecules. This is expected since the higher the interfacial pressure, the more compact the molecules are, and their mobility is reduced. In any case, Figures \ref{fig:NR} and \ref{fig:summ} show no evidence of multilayer formation within the duration of the experiment. This is not surprising since the spreading pressure of the DPPC monolayers is about $45$ mN/m \cite{Mansour2007,Hermans2014}. This conclusion cannot be obtained from the pressure-area isotherm alone.

\begin{table}[ht]
    \centering
    \begin{threeparttable}
    \caption{\label{tab:summ}Summary table with the two-layer model parameters at different interfacial pressures.}
    \begin{tabular}{lccc}
        \hline
        Parameter & $\Pi = 25$ mN/m & $\Pi = 35$ mN/m & $\Pi = 45$ mN/m \\ 
        \hline
        ${V_\mathrm{AC}}^*$ & 759 $\text{\AA}^3$ & 759 $\text{\AA}^3$ & 759 $\text{\AA}^3$ \\
        ${V_\mathrm{PC}}^*$ & 344 $\text{\AA}^3$ & 344 $\text{\AA}^3$ & 344 $\text{\AA}^3$ \\
        $t_\mathrm{AC}$ & (16.27 $\pm$ 0.01) $\text{\AA}$ & (17.09 $\pm$ 0.01) $\text{\AA}$ & (17.41 $\pm$ 0.01) $\text{\AA}$ \\
        ${t_\mathrm{PC}}^*$ & 9 $\text{\AA}$ & 9 $\text{\AA}$ & 9 $\text{\AA}$ \\
        $\sigma^*$ & 3.45 $\text{\AA}$ & 3.88 $\text{\AA}$ & 4.53 $\text{\AA}$ \\
        ${A_\mathrm{molec}}^\dag$ & (46.66 $\pm$ 0.02) $\text{\AA}^2$ & (44.42 $\pm$ 0.02) $\text{\AA}^2$ & (43.60 $\pm$ 0.02) $\text{\AA}^2$ \\
        $\phi_\mathrm{PC,w}$ & (18 $\pm$ 0.03)\% & (14 $\pm$ 0.03)\% & (12 $\pm$ 0.04)\% \\
        ${\mathrm{SLD}_{\mathrm{D_2O}}}^{\dag}$ ($\times10^{-6}$) & (5.812 $\pm$ 0.002) $\mathrm{\text{\AA}^{-2}}$ & (5.771 $\pm$ 0.003) $\mathrm{\text{\AA}^{-2}}$ & (5.660 $\pm$ 0.002) $\mathrm{\text{\AA}^{-2}}$ \\
        ${\mathrm{SLD}_{\mathrm{PC}}}^*$ ($\times10^{-6}$) & 1.74 $\mathrm{\text{\AA}^{-2}}$ & 1.74 $\mathrm{\text{\AA}^{-2}}$ & 1.74 $\mathrm{\text{\AA}^{-2}}$ \\
        ${\mathrm{SLD}_{\mathrm{AC}}}^*$ ($\times10^{-6}$) & 8.08 $\mathrm{\text{\AA}^{-2}}$ & 8.08 $\mathrm{\text{\AA}^{-2}}$ & 8.08 $\mathrm{\text{\AA}^{-2}}$ \\
        $\chi^2$ & 26.7 & 36.2 & 15.83 \\
        \hline
    \end{tabular}
    \begin{tablenotes}
        \item[*] Parameters taken from the literature \cite{Campbell2018a, Ocko1994}.
        \item[$\dag$] Treated as a free parameter in the fitting procedure.
    \end{tablenotes}
    \end{threeparttable}
\end{table}

\section{Conclusions}
\label{sec:concl}

We describe a new sample environment setup that allows one to perform \textit{in situ} simultaneous measurements of neutron reflectivity and interfacial rheology on the same sample. A rotational rheometer with DWR geometry has been coupled to a Langmuir trough that fits on the FIGARO anti-vibration table. An \textit{ad hoc} data acquisition programme has been developed to obtain and analyse torque and angular position signals that allow the calculation of interfacial dynamic moduli on-the-fly. Hence, the viscoelastic properties of fluid interfaces can be measured simultaneously with neutron reflectometry data. This combined facility allows studies on the interrelation between the microscopic structure and the mechanics of interfacial systems. 

We validated the performance of the full system by simultaneously studying the structural and rheological properties of DPPC monolayers at the air/water interface. Different aqueous subphases that yield different contrasts for NR have been used. The rheological behaviour of the samples has been studied by oscillatory measurements under frequency and strain sweeps. The NR results have been satisfactorily analysed with a simple monolayer model. The results yielded by simultaneous measurements using both interfacial shear rheometry and neutron reflectometry techniques agree well with previous results available in the literature. Collectively, the observed structural trends establish a direct molecular-scale foundation for the macroscopic rheological response: as the monolayer compresses, reduced molecular mobility and altered headgroup hydration lead to higher interfacial shear dynamic moduli.

The combination of instrumental techniques proposed here is especially suitable for studies such as: i) clarifying whether changes in the interfacial rheological properties are due to multilayer formation or not, ii) correlating the changes of dynamical and structural parameters in kinetic processes (adsorption, diffusion, etc.), iii) characterization of monolayer phase transitions, including thermodynamic, mechanical, and structural aspects, and iv) studies on interfacial systems where the molecules are not available in deuterated version but that show strong changes in their interfacial rheological properties: polymers, proteins/peptides in biological membranes, etc. This setup is essential to determine whether the data collected during a full-Q measurement (typically $1$ h duration: $5$-$15$ min at $\theta_1$ and $25$-$50$ min at $\theta_2$) correspond to a steady, transient, or out-of-equilibrium state. As a bonus, this combination of techniques allows for saving of experimental materials, which is of great interest when dealing with particularly expensive or precious samples. 

\section*{Acknowledgements}
The authors gratefully acknowledge Simon Wood for his support on mechanical design and fabrication, and the PSCM for access to its laboratories. We acknowledge the beam time allocation on FIGARO (\url{http://doi.ill.fr/10.5291/ILL-DATA.9-12-725}, \url{http://doi.ill.fr/10.5291/ILL-DATA.9-12-726}, \url{http://doi.ill.fr/10.5291/ILL-DATA.9-13-1089}). P.S.P. acknowledges the MICINN-ILL postdoc program for supporting his stay at the ILL. J.T. and M.A.R. acknowledge the support of the Spanish Ministerio de Ciencia e Innovaci\'on (MCIN) - Agencia Estatal de Investigaci\'on (MCIN/AEI/10.13039/501100011033) through the projects PID2020-117080RB-C54 and PID2023-147948OB-C33. A.M. acknowledges the financial support from MCIN/AEI/10.13039/501100011033 under grants PID2021-129054NA-I00 and PID2024-157988NB-I00, as well as the financial support from the Department of Education of the Basque Government under grant PIBA\_2023\_1\_0054 and from the IKUR Strategy under the collaboration agreement between Ikerbasque Foundation and Materials Physics Center.

\section*{Funding}
This work was supported by the Spanish Ministerio de Ciencia e Innovaci\'on (MCIN) - Agencia Estatal de Investigaci\'on (MCIN/AEI/10.13039/501100011033) through projects PID2020-117080RB-C54 and PID2023-147948OB-C33 (J.T. and M.A.R.).  
A.M. was supported by PID2021-129054NA-I00 funded by MICIU/AEI/10.13039/501100011033 and FEDER, UE; by the Department of Education of the Basque Government under grant PIBA-2023-1-0054; and by the IKUR Strategy project IKUR-Neutrónica 2025-2026 (NEU6.NANOBIO).  
P.S.P. acknowledges support from the MICINN–ILL postdoctoral program.

\section*{Conflicts of Interest}
The authors have no conflict of interest to disclose.

\section*{Data Availability}
Neutron Reflectivity data supporting the findings of this study are available from the ILL Data Portal (DOI: 10.5291/ILL-DATA.9-12-725, 10.5291/ILL-DATA.9-12-726 and 10.5291/ILL-DATA.9-13-1089). Data will be publicly available after the standard 3-year embargo period. Interfacial Shear Rheology and Langmuir data are available from the corresponding author upon reasonable request.

% \bibliographystyle{agsm}
% \bibliography{uu5023}

\end{document}